\documentclass[twocolumn,showpacs,preprintnumbers,nofootinbib,prd,aps,
superscriptaddress,10pt]{revtex4-1}

\usepackage{amsmath,amssymb}
\usepackage[normalem]{ulem}
\usepackage{textcomp}
\usepackage{hyperref}
\usepackage{bm}
\usepackage{graphicx}
\usepackage{psfrag}
\usepackage[usenames,dvipsnames]{xcolor}
\usepackage[utf8]{inputenc}
\usepackage[english, capitalise]{cleveref}
\crefname{chapter}{Chap.}{Chap.}
\crefname{section}{Sec.}{Sec.}
\Crefname{chapter}{Chapter}{Chapters}
\Crefname{section}{Section}{Sections}
\Crefname{eqs}{Eqs.}{Eqs.}		
\Crefformat{eqs}{Eqs.~(#2#1#3)}	

\definecolor{darkgreen}{rgb}{0,0.5,0}
\hypersetup{
	bookmarks=true,         
	unicode=false,          
	pdftoolbar=true,        
	pdfmenubar=true,        
	pdffitwindow=false,     
	pdfstartview={FitH},    
	pdftitle={Search for sub-solar mass compact binaries using the gravitational wave memory 
	effect},
	pdfauthor={M.~Ebersold, S.~Tiwari},
	pdfsubject={Gravitational Waves},
	pdfkeywords={gravitational waves, memory effect, general relativity},
	pdfnewwindow=true,      
	colorlinks=true,       
	linkcolor=red,          
	citecolor=cyan,        
	filecolor=magenta,      
	urlcolor=darkgreen,           
	linktocpage=true
}

\allowdisplaybreaks[4]

\newcommand{\bigO}{\mathcal{O}}

\def\no{\nonumber\\}
\newcommand{\Msun}{M_\odot}
\newcommand{\mem}{\textnormal{mem}}
\newcommand{\ri}{\mathrm{i}}

\graphicspath{{./Figures/}}

\begin{document}

\title{Search for nonlinear memory from subsolar mass compact binary mergers}

\date{\today}

\author{Michael Ebersold}
\affiliation{Physik-Institut, Universit\"at Z\"urich, Winterthurerstrasse 190, 8057 Z\"urich}

\author{Shubhanshu Tiwari}
\affiliation{Physik-Institut, Universit\"at Z\"urich, Winterthurerstrasse 190, 8057 Z\"urich}

\begin{abstract}
We present the first results of the search for nonlinear memory from subsolar mass binary black 
hole (BBH) mergers during the second observing run (O2) of the LIGO and Virgo detectors. The 
oscillatory chirp signal from the inspiral and merger of low mass BBHs ($M _\mathrm{Tot} \leq 0.4 
\Msun$) are at very high frequencies and fall outside the sensitivity band of the current 
ground-based detectors. However, the nonoscillatory memory signal during the merger saturates 
toward the lower frequencies and can be detected for those proposed BBHs. We show in this work that 
the morphology of the memory signal depends minimally upon the source parameters of the binary, 
thus only the overall amplitude of the signal changes and hence the result can be interpolated for 
extremely low mass BBH mergers. We did not find any signal which can be interpreted as a memory 
signal and we place upper limits on the rate of BBH mergers with $M _\mathrm{Tot} \leq 0.4 \Msun$ 
for the first time.
\end{abstract}

\pacs{
 04.30.-w, 
 04.30.Tv 
}

\maketitle
\section{Introduction} \label{sec:introduction}
Gravitational wave (GW) observations by Advanced LIGO~\cite{LIGO-2015} and Advanced 
Virgo~\cite{Virgo-2015} detectors have led to an unprecedented understanding of the population of 
compact binaries detectable by ground-based interferometers. In the first two observing runs ten
stellar mass binary black holes (BBHs) and one binary neutron star merger have been 
detected~\cite{LIGO_GW150914-2016,LIGO_BNS-2017,LIGO_Catalog-2019,LIGO_RatePop-2018}. 
No viable gravitational wave candidates with a component mass below $1 \Msun$ have been 
found~\cite{LIGO_subsolar-2018,LIGO_subsolar-2019}. In this paper, we present a novel approach 
to search for subsolar mass compact objects by making use of the GW memory effect.

Gravitational waves are usually thought of as purely oscillatory perturbations propagating on the 
background metric at the speed of light. However, all GW sources are subject to the GW memory 
effect, which manifests in a difference of the observed GW amplitudes at late and early times.
In an ideal, freely falling GW detector, the GW memory causes a permanent displacement after the GW
has passed. Here, we focus on the nonlinear memory, also called 
``Christodoulou memory"~\cite{christodoulou-1991,wiseman-1991,thorne-1992,blanchet-1992} (``memory" 
henceforth). It arises from GWs sourced by previously emitted GWs and is therefore directly related 
to the nonlinearity of general relativity (GR). The memory is present in all GW sources since 
it is not produced directly by the source but rather by its radiation. Although the memory has not 
been observed yet, the prospects are looking good that this will happen in the near 
future~\cite{cordes-2012,favata-2009-2,lasky-2016,johnson-2019,divakarla-2019,huebner-2020,boersma-2020}.
 From a more theoretical perspective, the memory effect and its variants can be interpreted in 
terms of conserved charges at null infinity and ``soft 
theorems"~\cite{strominger-2016,pasterski-2016,zhang-2017}. 

The amplitude of a GW memory signal from a compact binary merger is monotonically increasing, very 
slowly during the inspiral; then, it jumps during the merger and finally saturates at its final 
value over the ringdown. This jump during the merger manifests as a burst signal; its duration 
depends upon the chirp mass of the binary system. Lighter systems have shorter burst duration. 
If the timescale of the burst is short compared to the inverse frequency of the detector's 
sensitive band, the memory signal can be approximated by a step function with an amplitude spectral 
density proportional to $1/f$, $f$ being the frequency. Therefore, the memory of a high-frequency 
burst leads to a low-frequency component coming from arbitrarily short 
bursts~\cite{braginski-1987}. Since the oscillatory signal from a subsolar mass compact binary 
mergers is well above LIGO's sensitive band, the memory burst is an example of ``orphan memory" as 
memory signals with no detectable parent were called in Ref.~\cite{mcneill-2017}.

Subsolar mass compact objects have never been observed, and there exists no mechanism in 
conventional stellar evolution models to form them. Black holes are supposed to be heavier than the 
Chandrasekhar limit of approximately $1.4 \Msun$, set by the proton mass~\cite{chandrasekhar-1931}, 
and neutron stars are expected to have masses above $0.9 \Msun$~\cite{strobel-2001,kaper-2006}. 
However, there exist several ideas for how subsolar mass compact objects could form. Some proposals 
link the existence of such objects to dark matter.
It has for example been suggested that cosmologically significant numbers of black holes could form 
out of vacuum bubbles that nucleated during inflation and collapsed after inflation 
ended~\cite{deng-2017}. Other proposals suggest that during a first-order QCD phase transition in 
the radiation era large primordial overdensities on the scale of Hubble volume at that time would 
suddenly collapse. The abundance and mass distribution of any such primordial black holes depend on 
the equation of state of the early Universe and the spectrum of primordial 
inhomogeneities~\cite{jedamzik-1997,cardall-1998,sobrinho-2016,byrnes-2018,carr-2019}. 
The existence of subsolar mass compact objects would be a smoking gun for a primordial 
origin, and they could arguably constitute a significant fraction of the cold dark matter 
density. Alternative possibilities include dark matter particles interacting with nuclear matter 
inside neutron stars leading to their collapse~\cite{bramante-2014,kouvaris-2018} or the existence 
of subsolar mass binary black holes formed out of dark matter particles~\cite{shandera-2018}.\\

\section{Memory waveform from BBH mergers} \label{sec:memory}

Numerical Relativity (NR) waveforms of binary black hole mergers usually do not contain memory. 
This is because it is generally difficult to extract a nonoscillatory or direct current (DC)
component from NR data and highly depends on the extraction 
method~\cite{pollney-2011,taylor-2013,bishop-2016,blackman-2017}. However, having 
the oscillatory waveform, the memory contribution can be computed separately using inputs 
from Refs.~\cite{favata-2009,favata-2010,talbot-2018,ebersold-2019}. 

It is convenient to decompose the GW polarizations into spin-weighted spherical harmonic modes 
$h^{\ell m}$ via
\begin{align}
	h_+ - \ri h_\times = \sum_{\ell = 2}^{\infty} \sum_{m = -\ell}^{\ell} h^{\ell m} Y^{\ell 
	m}_{-2}(\iota, \Phi)\,,
\end{align}
where the angles $\iota$ and $\Phi$ denote inclination and a reference phase. We use the same 
conventions on the polarizations and modes as in Ref.~\cite{boetzel-2019}.
The memory contribution to the $h^{\ell m}$-modes can be computed from the oscillatory modes by
\begin{align}\label{eq:hlmmem}
	h^{\ell m}_\mem =&\; -\frac{R}{c} \sqrt{\frac{(\ell-2)!}{(\ell+2)!}}
		\sum_{\ell'=2}^{\infty} \sum_{\ell''=2}^{\infty} \sum_{m'=-\ell'}^{\ell'}
		\sum_{m''=-\ell''}^{\ell''} \no
	&\times G^{\ell \ell' \ell''}_{m m'm''} \int_{-\infty}^{T_R} d t \,\dot{h}^{\ell'm'} 	
		\dot{\bar h}^{\ell''m''}\,,
\end{align}
where $R$ is the distance to the binary and $G^{\ell \ell' \ell''}_{m m'm''}$ is an angular 
integral of a product of three spin-weighted spherical harmonics which imposes some selection 
rules. Explicitly, it is given by
\begin{align}\label{eq:G}
	G^{\ell \ell' \ell''}_{m m' m''} =&\; \int d \Omega' \, \bar Y{^{\ell m}}(\Omega') \, 
		Y^{\ell' m'}_{-2}(\Omega') \, \bar Y^{\ell'' m''}_{-2} (\Omega') \no
	=&\; (-1)^{m + m'} \sqrt{\frac{(2\ell +1)(2\ell' +1)(2\ell'' +1)}{4\pi}} \no
	&\times 
		\begin{pmatrix}
			\ell & \ell' & \ell'' \\
			0 & 2 & -2
		\end{pmatrix}
		\begin{pmatrix}
			\ell & \ell' & \ell'' \\
			-m & m' & -m''
		\end{pmatrix} \,,
\end{align}
where the angles $\Omega'$ describe a sphere centered at the source and the brackets denote the 
Wigner 3-$j$ symbols. The dominant memory mode turns out to be $h^{20}_\mem$, and it primarily 
contributes to the $h_+$-polarization for nonprecessing equal mass binaries. The memory is mainly 
sourced by the dominant oscillatory modes $\ell = |m| = 2$. Including additional higher order modes 
in the calculation of the memory leads to $\bigO(10\%) $ change in the memory amplitude. 
Precessing and unequal mass systems lose memory amplitude in $h_+$ but also source memory in the 
$h_\times$-polarization~\cite{talbot-2018}.
\begin{figure}
	\includegraphics[scale=0.56]{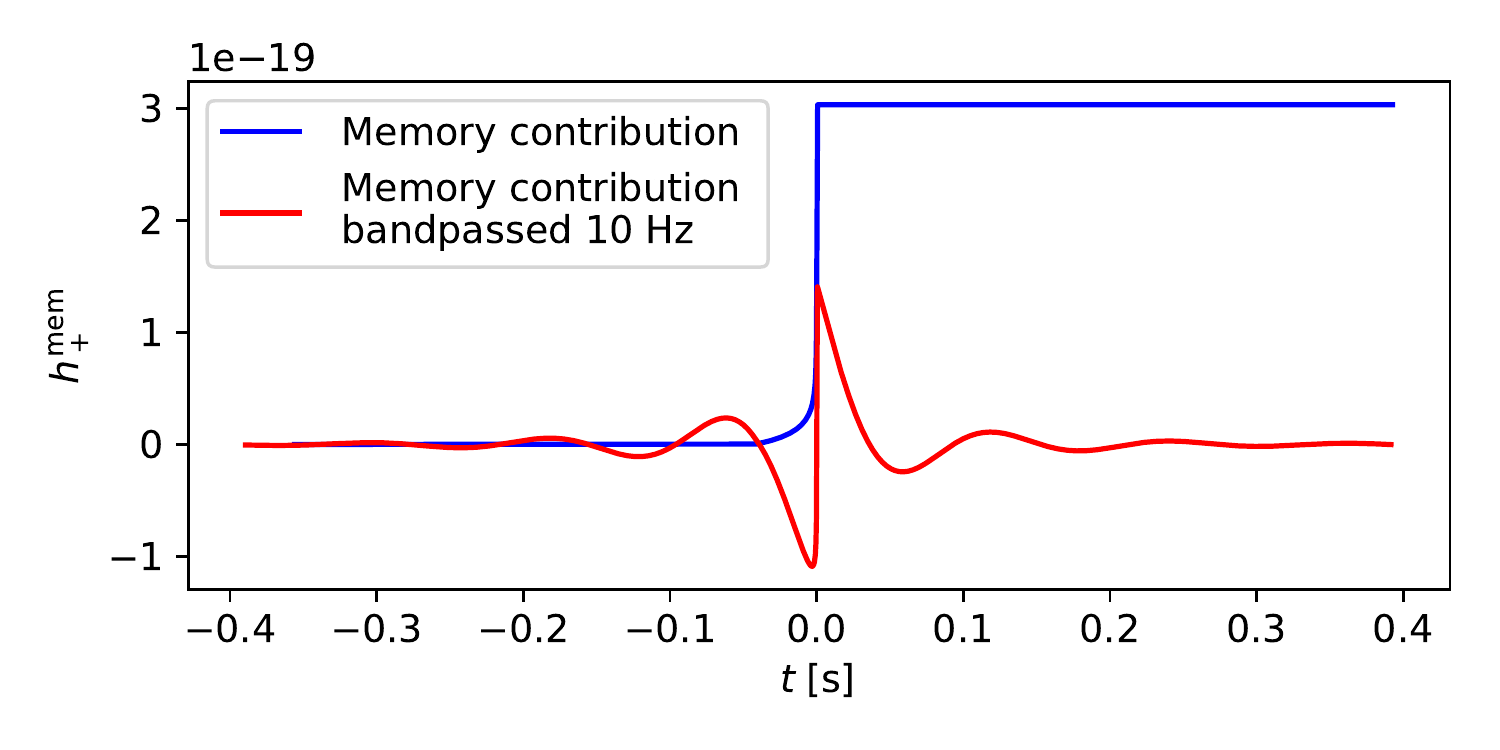}
	\caption{The time-domain memory waveform of an edge-on binary black hole merger of $2 M_\odot$ 
	total mass at 10 kpc. The black holes are of the same mass and nonspinning. Note that we 
	neglect any memory from the inspiral and only consider the memory generated from about -0.04 s 
	until the merger.}
	\label{fig:memwaveform}
\end{figure}
\begin{figure}
	\includegraphics[scale=0.3]{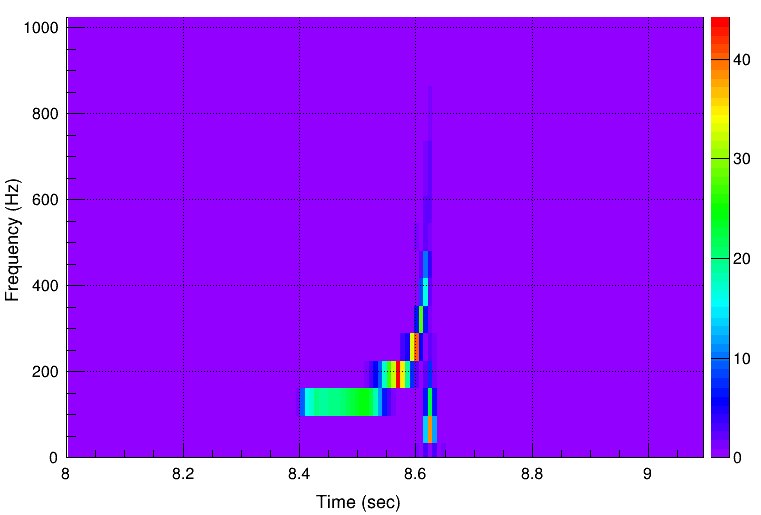}
	\caption{Whitened spectrogram with O2 noise power spectral density of the GW signal of an equal 
	mass binary black hole merger at SNR$\approx 100$. One can clearly distinguish the chirp signal 
	from the memory burst saturating toward lower-frequency cutoff of the analysis at the time of 
	the merger. This signal is from a 5-5 $\Msun$ system; the oscillatory part is shown above 100 
	Hz, and the memory contribution is slightly exaggerated. For lower mass binaries, the chirp 
	signal will rise in frequency, whereas the memory signal will always look the same in a 
	time-frequency representation and only change in amplitude.}
	\label{fig:spectogram}
\end{figure}
We are only interested in the burst of memory during the merger. To this end we use the oscillatory 
waveform modes from the NR surrogate waveform model ``NRSur7dq4"~\cite{varma-2019} to compute 
memory from. This waveform model contains all modes $\ell \leq 4$ and can handle generically 
spinning binaries up to a mass ratio $m_1/m_2 = 4$ and spin magnitudes $\chi_1, \chi_2 \leq 0.8$. 
Using~\cref{eq:hlmmem}, we compute the memory strain from the surrogate waveform by numerical 
integration. The lower-frequency cutoff of current ground-based interferometers is around 10 Hz; 
therefore, we cut off the frequencies below 10 Hz of the memory signals with a high-pass filter. A 
memory waveform from a BBH merger with and without a high-pass filter is shown 
in~\cref{fig:memwaveform}. One can clearly see that the bandpassed memory signal appears like a 
short-duration burst. The oscillations around the central peak are merely an artifact of 
bandpassing. This burst can be detected by interferometric detectors.  An example of how the GW 
detectors see the memory is given in~\cref{fig:spectogram}. For illustrative purposes, we consider 
for this representation a $5-5 \Msun$ BBH system with enhanced memory content in order to clearly 
show the oscillatory part and the memory part of the signal. The latter has its peak value at 
around 100 Hz as it is the most sensitive part of the detector. This does not change for lower mass 
BBHs, as one can also see in~\cref{fig:memfreq}, whereas the oscillatory part moves toward higher 
frequencies and eventually beyond the sensitivity of LIGO and Virgo.
\begin{figure}
	\includegraphics[scale=0.53]{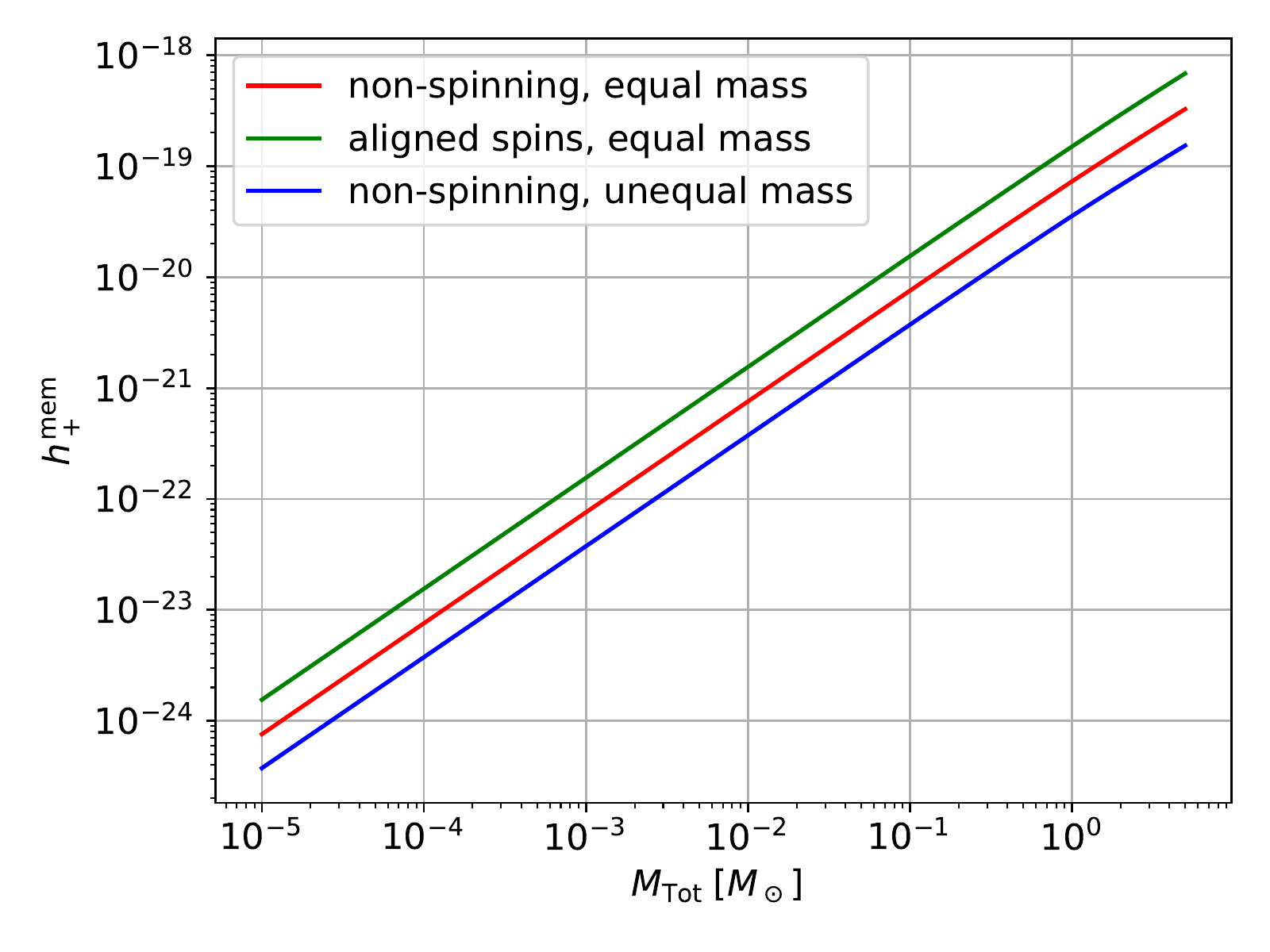}
	\caption{The maximum of the bandpassed (10 Hz) memory amplitude as a function of the total mass 
	for an edge-on binary at 10 kpc. In the aligned spin case, both black holes are spinning with 
	spin magnitude of 0.8, and the unequal mass system has a mass ratio of $m_1/m_2 = 3$. The 
	amplitude scales linearly well below $1 \Msun$.}
	\label{fig:massamp}
\end{figure}

The morphology of the waveform from BBH systems with same spins and mass ratio is the same; only 
the amplitude and the time axis scale proportional to the total mass. The same is true for the 
memory waveform. In~\cref{fig:massamp}, we plot the memory amplitude vs the total mass of a binary 
system for a nonspinning equal mass and a $m_1/m_2 = 3$ system as well as for an aligned spin 
equal mass system. The linear behavior with total mass is clearly seen, although for large masses, 
we lose amplitude due to the low-frequency cutoff. One can think of the memory burst not happening 
fast enough for systems above $1 \Msun$. This behavior also explains why with our search method we 
cannot see the memory from the detected stellar mass BBHs. The memory waveform depends differently 
on the inclination angle $\iota$ between the binaries orbital angular momentum axis and the line of 
sight toward the observer than the oscillatory waveform. The dominant memory in the plus 
polarization scales like $h_+^\mem \sim \sin^2 \iota (17+\cos^2 \iota)$~\cite{favata-2010}. This 
dependence holds well when including higher modes for equal mass, aligned spin systems, but would 
get more complex for unequal mass and/or precessing systems.\\

\section{Search for memory signals in LIGO data} \label{sec:search}
The data used in this study is part of the O2 Data Release through the Gravitational Wave Open 
Science Center~\cite{LIGOData}. Our dataset is the second observing run of LIGO and Virgo 
detectors. In this work, we have used the data from the two LIGO detectors located in Livingston 
and Hanford, which range from November 30, 2016, to August 25, 2017. The Advanced Virgo detector 
was less sensitive than the Advanced LIGO detectors, with a binary neutron star range that was 
roughly a factor of 2–3 lower. As a result of this, including the Virgo dataset did not improve the 
sensitivity to the short-duration searches presented in this paper. We thus present the analysis of 
only the Hanford-Livingston data.

Over the course of O2, the live time of the data collected by the two LIGO detectors was about 158 
days for Hanford and about 154 days for Livingston. The amount of coincident data between the two 
detectors is approximately 118 days. The analysis that is in this work is performed by dividing the 
run into reduced periods of consecutive time epochs (called “chunks”). Each chunk is composed of 
about 5 days of live time, resulting in 21 chunks in total. Performing the analyses in chunks takes 
into account nonstationary noise levels of the detectors over the duration of the observing run.  

To interpret the detection sensitivity of the memory signals, we have used the unmodeled search 
\textit{coherent WaveBursts} (cWB)~\cite{klimenko-2008,klimenko-2016}. cWB is an algorithm based on 
the maximum-likelihood-ratio statistic applied to power excesses in the time-frequency domain. This 
analysis is done by using a wavelet transform at various resolutions, so as to adapt the 
time-frequency characterization to the signal features.
The search setting and configuration for this work are exactly the same as reported in the all-sky 
search for short-duration transients during the second observing run~\cite{all-skyo2}. As the 
spectral content of memory signal falls in the low-frequency regime, we have used only the 
low-frequency bin of the analysis. No further tunings are done for better detecting the memory 
signals, but it should be noted that the cWB pipeline can be tuned for the memory signals; this 
will be presented in future works.

The low-frequency analysis of cWB covers the parameter space ranging from 32–1024 Hz, and performs 
a down sampling of the data. The triggers are divided into two different bins. The first bin, LF1, 
is polluted by nonstationary power spectrum lines and a class of low-frequency, short-duration 
glitches known as “blip” glitches for which there is no specific data quality veto~\cite{detchar}. 
These are selected using the same criteria as described in Ref.~\cite{unmod-150914}: nonstationary 
lines localize more than 80\% of their energy in a frequency bandwidth of less than 5 Hz; blip 
glitches are identified according to their waveform properties so that their quality factor (Q) is 
less than 3. The second bin, LF2, contains the remaining low-frequency triggers. Unfortunately, the 
morphology of the subsolar mass memory signal is such that it falls in the LF1 bin lowering the 
sensitivity of the search. A better discriminant of the blip glitches can potentially improve the 
sensitivity of the search for these signals.  

As mentioned in the previous section, the full set of coincident data is divided into 21 chunks. 
The background distribution of triggers for each individual chunk is calculated by time shifting 
the data of one detector with respect to the other detector by an amount that breaks any 
correlation between detectors for a real signal. Each chunk was time shifted to give about 500 
years of background data, which allows the search to reach the statistical significance of 1 per 
year while allowing for a trial factor of 2 for the two bins in the low-frequency analysis.
As reported in Ref.~\cite{all-skyo2}, the search used here for the memory signals from subsolar 
mass BBH does not find any new events apart from a subset of known BBH signals already found and 
detailed in Ref.~\cite{LIGO_Catalog-2019}. We have used this result of null detection to put upper 
limits on the rates of subsolar mass BBH mergers using memory.

\begin{figure}
	\begin{minipage}{0.45\textwidth}
		\includegraphics[scale=0.42]{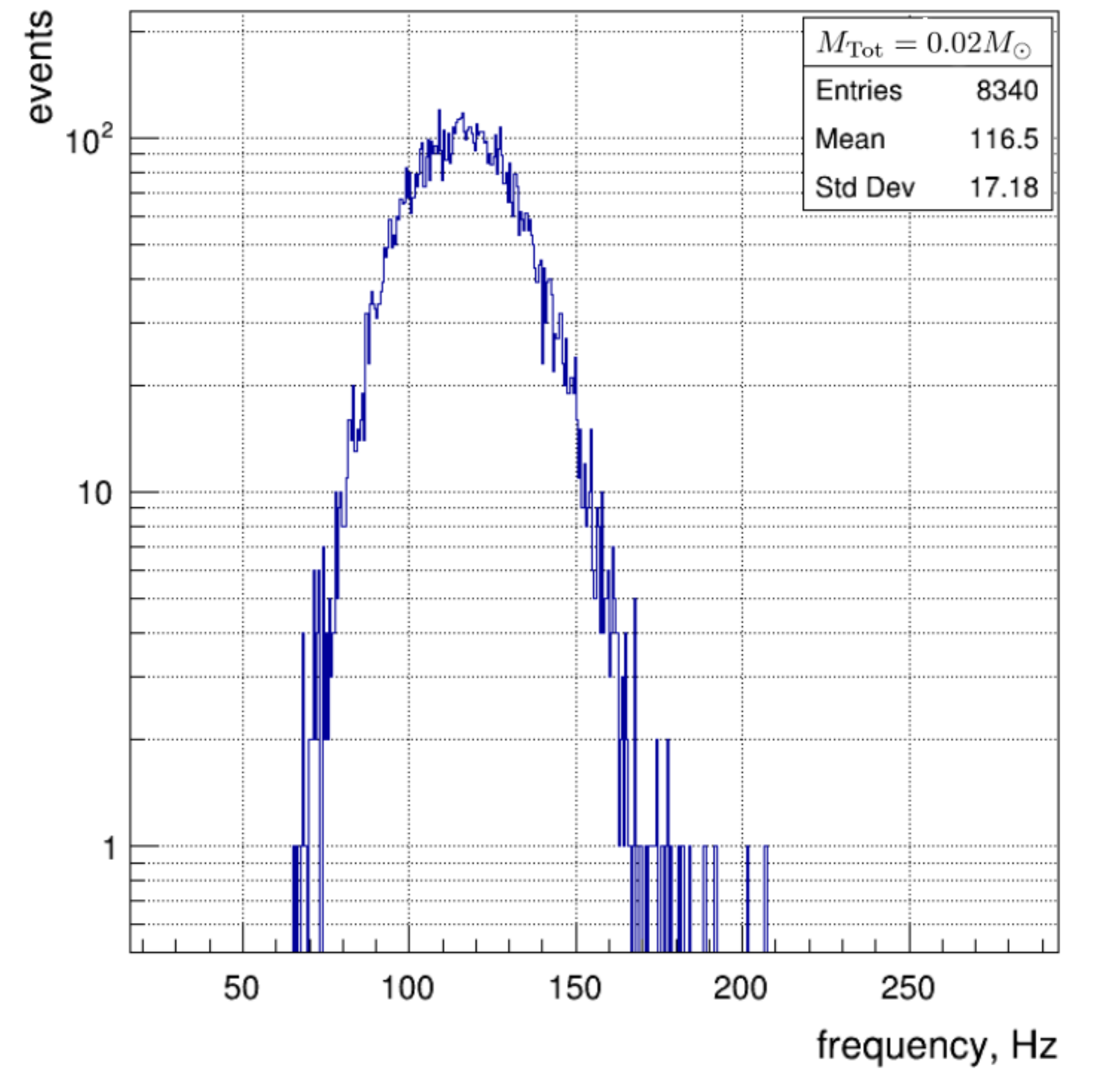}
	\end{minipage}
	\caption{Histogram of the central frequency of recovered memory injections for a $0.02 \Msun$ 
	system. In fact, the memory from low mass systems always saturates around the detectors' most 
	sensitive region, which is around 100 Hz.}
	\label{fig:memfreq}
\end{figure}

To study the sensitivity of our search, we injected six different memory signal types into the 
detector data and searched for them using cWB. We injected memory signals from equal mass binaries 
with total masses of $0.02 \Msun$, $0.2 \Msun$, and $2 \Msun$, once nonspinning and once with an 
aligned spin of 0.8 spin magnitude. Subsolar mass compact objects of primordial origin are often 
assumed to have near zero spin~\cite{deluca-2019,mirbabayi-2019}, although there are also proposals 
of almost maximally spinning primordial black holes~\cite{harada-2017}. All injections are 
uniformly distributed in sky direction, and distance distribution is uniform in volume. 
In~\cref{fig:memfreq}, we show the reconstructed central frequency distribution for a $0.02 \Msun$ 
system. The spectral property of the memory signal from various subsolar mass systems looks similar 
and does not depend on the total mass. Moreover, we show through injections that the visible range 
is linear in total mass and only the overall amplitude matters (see~\cref{fig:massrange}). 
Therefore, this result can be easily extrapolated to very low mass systems.\\
\section{Results} \label{sec:results}
The sensitivity of our search for subsolar mass memory is characterized by its range: the distance 
within which a memory signal could be detected with an inverse false alarm rate (iFAR) of greater 
than or equal to 1 yr. 
\begin{figure}
	\includegraphics[scale=0.59]{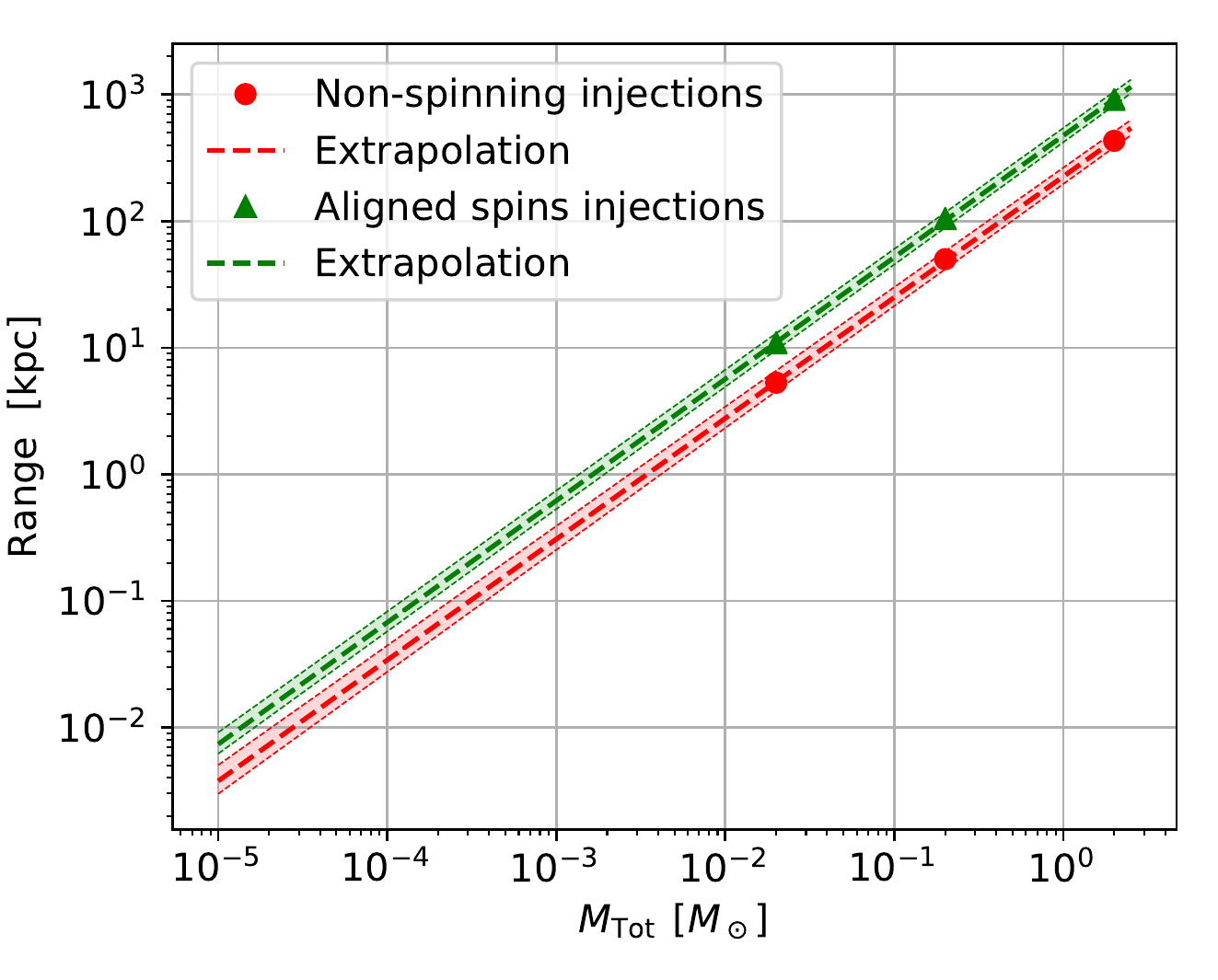}
	\caption{Range of the cWB analysis to recover a memory signal from subsolar mass BBH mergers at 
	an iFAR $\geq$ 1 yr as a function of total mass. Only equal mass coalescences are considered. 
	The result from the injections can be extrapolated to very small masses because the search 
	sensitivity depends only on the amplitude of the memory signal. The shaded regions show the $1 
	\sigma$ uncertainties.}
	\label{fig:massrange}
\end{figure}
We compute the sensitive volume, which is the sphere built with radius being the range. Since we 
are looking for signals from distances less than 1 Mpc, well below redshift becomes important, we 
compute the sensitive volume according to Ref.~\cite{tiwari-2018} in the limit $z \rightarrow 0$,
\begin{align}
	\langle V \rangle = 4 \pi \int d\theta \, dr \, r^2 \, p_\mathrm{pop}(\theta)\, f(r,\theta)\,,
\end{align}
where $p_\mathrm{pop}(\theta)$ is the distribution function for the astrophysical population and 
$f(r,\theta)$ is the detection efficiency measuring the probability of recovering a signal with 
parameters $\theta$ at distance $r$. It is obtained by signal injections into O2 data. Since we 
have injected signals with fixed inclination angle of $\iota = \frac{\pi}{2}$, we account for the 
random distribution by multiplying the detection efficiency with the average value of the 
inclination angle dependence ($\simeq 0.51$). Figure~\ref{fig:massrange} shows the range of the 
search for nonspinning and aligned spins systems of different total masses.

We can compute an upper limit of the merger rate $R_i$ of our populations of subsolar mass black 
holes. Assuming that the observation of a GW signal follows a Poisson process, the rate limit is 
inversely proportional to the sensitive volume of the population. We estimate the upper limit on 
the binary merger rate to $90 \%$ confidence level at iFAR greater than or equal to 1 yr by
\begin{align}
	R_i = \frac{2.3}{\langle V_{\mathrm{iFAR} =1 \mathrm{yr}} T\rangle_i}\,,
\end{align}
where $T$ denotes the length of coincident detector data, which after data quality cuts amounts to 
114.78 days.
\begin{figure}
	\includegraphics[scale=0.66]{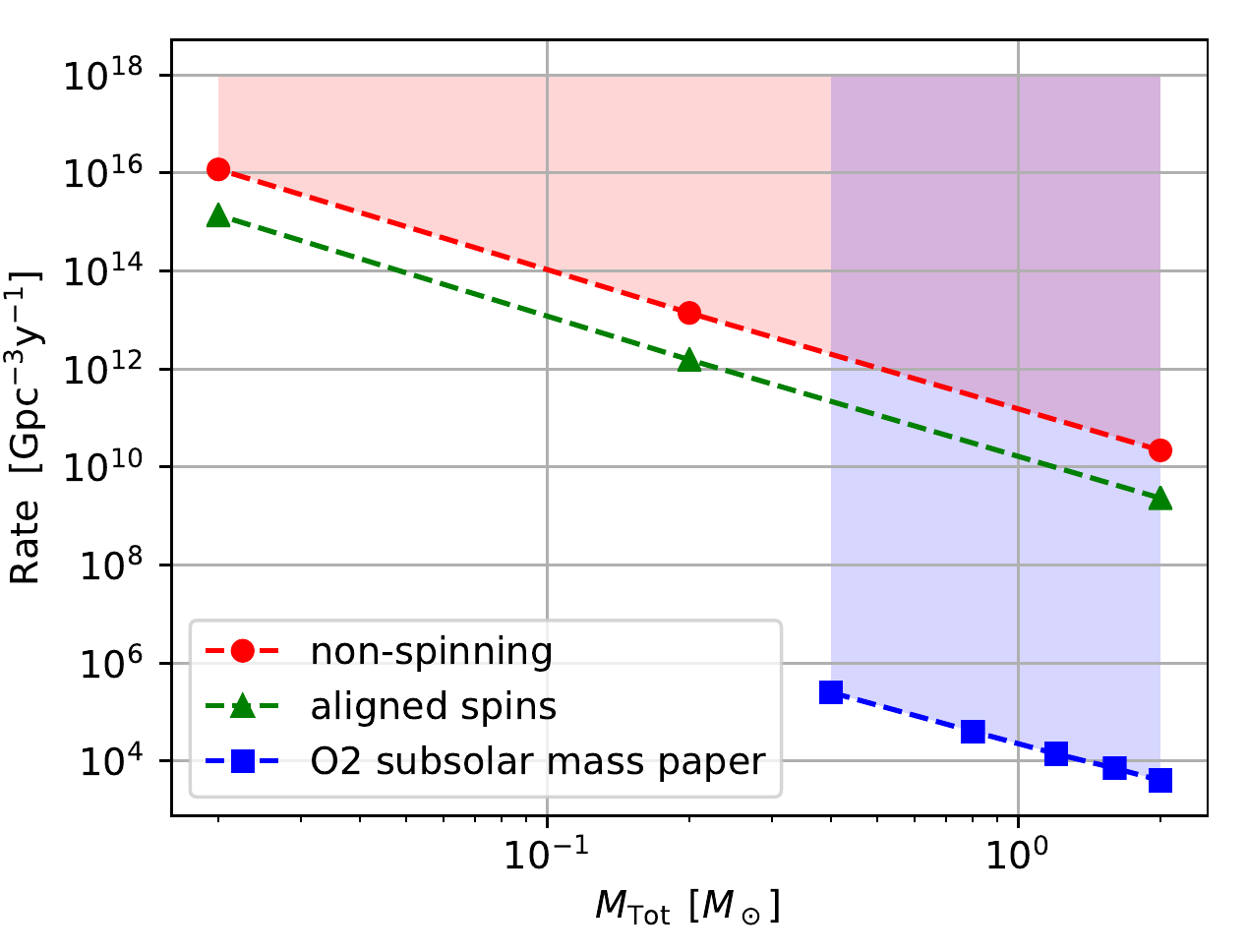}
	\caption{The constraint on the merger rate density for equal mass binaries. Shaded regions 
	represent excluded values. We can compare the constraint on nonspinning subsolar mass black 
	hole mergers to the results obtainted by LIGO using a matched-filtering search in the O2
	subsolar mass paper~\cite{LIGO_subsolar-2019}. For completeness, we also show the constraint on 
	an aligned spin population.}
	\label{fig:rate}
\end{figure}
Figure \ref{fig:rate} shows our rate upper limits and the results from LIGO obtained in 
Ref.~\cite{LIGO_subsolar-2018}. By comparing, one recognizes immediately that our rate upper limits 
for BBH systems greater than or equal to 0.4 $\Msun$ are several orders of magnitude worse. This is 
to be expected since the memory signal is itself about an order of magnitude weaker than the 
oscillatory signal. Moreover, the dependence of the memory on the inclination angle leads to a 
further loss. However, for lower masses, the detection of the oscillatory signal becomes more 
difficult, and eventually it will fall out of the detectors sensitivity band. In contrast, the 
memory contribution can be detected for arbitrarily low mass compact binary mergers.\\

\section{Conclusion} \label{sec:conclusion}
In this paper, we have employed the all-sky search for generic GW transients for the detection of 
memory signal from subsolar mass BBH mergers. Even though our constrains are not competitive for 
the regions of the parameter space where template-based searches for the oscillatory part of the 
subsolar mass BBH are done, searching for a memory only part of the signal has distinctive 
advantages. A memory only search can drastically increase the parameter space of the search, as it 
can cover very low mass regions and also highly spinning systems. It should be noted that the 
memory signal is not as energetic as the oscillatory signal as apparent with our upper limits but 
going to lower masses for the template based searches for the oscillatory signal will be 
computationally infeasible with little to no gain.

Due to the lack of dependence of spectral and morphological features on the source parameters 
(mass ratio, spins, etc.) of the memory signal, it will be challenging to precisely estimate the 
source parameters with memory only detection. But it should be noted that if a memory only signal 
is detected without the detection of the oscillatory signal we can conclude that the signal arises 
from a system which is beyond the parameter space covered by the template-based searches.

Further improvements for the detection of memory only signal can be made for instance with a highly 
tuned search for the detection of these sources and a better understanding of the very short noise 
transients known as blips. Moreover, a population of subsolar mass mergers can have a stochastic 
background which might be detectable by the current or future generation of detectors.\\

\section{Acknowledgements}
We thank Maria Haney and Giovanni Prodi for insightful discussions and comments and Colm Talbot for 
an early review. M.~E. and S.~T. are supported by the Swiss National Science Foundation and a 
Forschungskredit of the University of Zurich. This research has made use of data, software, and/or 
web tools obtained from the Gravitational Wave Open Science Center 
(https://www.gw-openscience.org), a service of LIGO Laboratory, the LIGO Scientific Collaboration, 
and the Virgo Collaboration. LIGO is funded by the U.S. National Science Foundation. Virgo is 
funded by the French Centre National de Recherche Scientifique (CNRS), the Italian Istituto 
Nazionale della Fisica Nucleare (INFN), and the Dutch Nikhef, with contributions by Polish and 
Hungarian institutes. The authors gratefully acknowledge the support of the NSF CIT cluster for the 
provision of computational resources.

\bibliographystyle{apsrev4-1}
\bibliography{references}

\end{document}